\documentclass[manuscript]{aastex} %\documentclass[preprint]{aastex}
\usepackage{float}
\usepackage{color}
\usepackage{amsmath}
\usepackage[flushleft]{threeparttable}
\usepackage{url}

\begin{document}

\title{The Search for a Complex Molecule in a Selected Hot Core Region: A Rigorous Attempt to Confirm trans-Ethyl Methyl Ether toward W51 e1/e2}

\author{P. Brandon Carroll, Brett A. McGuire}
\affil{Division of Chemistry and Chemical Engineering, California Institute of Technology, 1200 E. California Blvd Pasadena CA, 91125, USA}
\author{Geoffrey A. Blake}
\affil{Division of Chemistry and Chemical Engineering and Division Geological and Planetary Sciences, California Institute of Technology, 1200 E. California Blvd Pasadena CA, 91125, USA}
\author{A. J. Apponi, L. M. Ziuyrs}
\affil{Departments of Chemistry and Astronomy, University of Arizona, 933 North Cherry Avenue, Tucson, AZ 85721, USA}
\author{Anthony Remijan}
\affil{National Radio Astronomy Observatory, 520 Edgemont Road, Charlottesville VA, 22901-2475, USA}

\begin{abstract}
An extensive search has been conducted to confirm transitions of \textit{trans}-ethyl methyl ether (tEME, C$_2$H$_5$OCH$_3$), toward the high mass star forming region W51 e1/e2 using the 12 m Telescope of the Arizona Radio Observatory (ARO) at wavelengths from  2 mm and 3 mm. In short, we cannot confirm the detection of tEME toward W51 e1/e2 and our results call into question the initial identification of this species by \citet{FuchsSpace}. Additionally, reevaluation of the data from the original detection indicates that tEME is not present toward W51 e1/e2 in the abundance reported by Fuchs and colleagues. Typical peak-to-peak noise levels for the present observations of W51 e1/e2 were between 10 - 30 mK, yielding an upper limit of the tEME column density of $\leq$ 1.5 $\times$ 10$^{15}$ cm$^{-2}$. This would make tEME at least a factor 2 times less abundant than dimethyl ether (CH$_3$OCH$_3$) toward W51 e1/e2. We also performed an extensive search for this species toward the high mass star forming region Sgr B2(N-LMH) with the NRAO 100 m Green Bank Telescope (GBT). No transitions of tEME were detected and we were able to set an upper limit to the tEME column density of $\leq$ 4 $\times$ 10$^{14}$ cm$^{-2}$ toward this source. Thus, we are able to show that tEME is not a new molecular component of the interstellar medium and that an exacting assessment must be carried out when assigning transitions of new molecular species to astronomical spectra to support the identification of large organic interstellar molecules.
\end{abstract}

%\citet{Halfen1}, \citet{Hollis2}, \citet{Hollis1}

\section{Introduction}
It is quite clear that our understanding of the molecular complexity of interstellar and circumstellar environments is rapidly growing. It is also apparent that our understanding of interstellar molecular synthesis is presently incomplete; observations of new interstellar molecules are currently outpacing the model predictions as to how these interstellar species are formed in astronomical environments. In addition, many searches for interstellar species have focused on complex organic molecules of biological significance, (e.g. \citet{Zaleski}, \citet{Loomis}, \citet{Belloche}). Since the detection of glycolaldehyde (HOCH$_2$CHO) there have been $\sim$60 new molecular species claimed in interstellar and circumstellar environments. Furthermore, a majority of these claimed detections have involved complex organic species including alcohols (vinyl alcohol (CH$_2$CHOH), \citet{Vinyl}; ethylene glycol (HOCH$_2$CH$_2$OH), \citet{Ethylene}); aldehydes (propenal (CH$_2$CHCHO), propanal (CH$_3$CH$_2$CHO), \citet{Propenal}); amino acids (glycine (NH$_2$CH$_2$COOH), \citet{Kuan}) sugars (dihydroxyacetone (CH$_2$OHCOCH$_2$OH),  \citet{Widicus}) and ethers (C$_2$H$_5$OCH$_3$, hereafter tEME, \citet{FuchsSpace}). 

Large organic molecules typically have high line strength (S$_{ij}\mu^2 \geq$ 50 D$^2$), low energy transitions ($\leq$ 50 K) that span the millimeter and submillimeter spectrum (e.g. \citet{Carroll}). Thus, it appears that the unambiguous identification of large molecules would be straightforward given the number of transitions available to search. Yet, the detection of new molecules becomes difficult at millimeter and submillimeter wavelengths due in large part to the line confusion of more well-known interstellar species, including isotopic variants. It has been estimated that in the 2 mm and 3 mm windows, there are approximately 10 lines per 100 MHz at sensitivity levels of 10 mK, toward high mass hot molecular cores (HMCs) \citep{Halfen1}. In the case of Sgr B2(N-LMH), perhaps the most well studied region to search for new interstellar species, the chance of finding a line at a particular LSR velocity ($\pm$ 2 km s$^{-1}$) of a measured spectral line frequency is $\sim$40\%, assuming simple Gaussian line profiles \citep{Halfen1}. Searching a less complicated source than Sgr B2(N-LMH) can partially mitigate this obstacle; however, the problem of coincident spectral features interfering with the detection of a new interstellar molecule still persists toward any chemically rich source.

The challenges in the identification of a new interstellar species have been reported by \cite{Snyder1}. The authors suggest ways to overcome these challenges by assigning a set of criteria that must be met before the identification of a new interstellar molecules is confirmed. These criteria can be summarized as follows: 1) The transition frequencies searched for must be accurate to within a few kHz. In addition, the most favorable transitions to search for are multiply degenerate (if possible), high line strength, and low energy. The criteria of high line strength and low energy depends on the molecule. 2) The LSR velocities between transitions must be consistent. 3) If possible, the transitions of a new molecular species must be separated by any interfering features by the Rayleigh criterion in order to claim that transition. 4) The relative intensities between transitions must be consistent with radiative transfer based on the physical conditions of the region. Finally 5), if possible, connecting transitions at higher and lower quantum numbers to the claimed transition should be detected. These criteria were applied to the claimed detections of glycine (\citealt{Kuan}) and dihydroxyacetone (\citealt{Widicus}) and both of the claimed detections were rejected (\citealt{Snyder1} and \citealt{Apponi1}, respectively). Conversely, the criteria were utilized to confirm the detection of glycolaldehyde (\citealt{Halfen1}) toward Sgr B2(N-LMH) at the 99.8\% confidence level. As demonstrated by \citet{Snyder1} and \citet{Apponi1}, the detection of a large organic molecule based on even 10 to 20 transitions can be tenuous.

In 2005, an extensive survey was performed by Fuchs and colleagues to search for interstellar trans-ethyl methyl ether, C$_2$H$_5$OCH$_3$, toward several high mass HMCs \citep{FuchsSpace}. This work was motivated by the previously reported observation of a single tEME transition towards Orion KL and W51 e1/e2 (\citealt{Charnley}). As a result of their survey, a detection of tEME was claimed toward the high mass star forming region W51 e2. %with a column density of 2 $\times$ 10$^{14}$ cm$^{-2}$ and a rotational temperature of $\sim$ 70 K. 

%W51 is a complex of HII regions located $\sim$ 7 kpc away in the Sagittarius spiral arm. The W51 e1 and e2 regions have been primary targets in surveys for large molecules e.g. \citet{Kalenskii}. At 2 mm, \citet{Zhang1} imaged dimethyl ether (CH$_3$OCH$_3$), methyl formate (HCOOCH$_3$) and methyl cyanide (CH$_3$CN). At 3 mm wavelengths, \citet{Snyder2} imaged four formic acid (HCOOH) transitions and detected transitions of ethyl cyanide (C$_2$H$_5$CN) and HCOOCH$_3$. Finally, \citet{Remijan1}  measured the density and temperature of e1 and e2 using CH$_3$CN. From that survey, they found a temperature of T$_{ex}$ = 153(21) for W51 e2 and T$_{ex}$  = 123(11) for W51 e1 and, further, that the chemistry occurring in the W51 e1/e2 region may be conducive to the formation of yet larger organic molecules.

This would make tEME the fourth largest molecule to be detected in the interstellar medium (ISM). The three molecules larger than tEME, HC$_{11}$N, C$_{60}$, and C$_{70}$, posses symmetry that greatly facilitates their detection. However, tEME lacks such symmetry. Determination of the tEME abundance therefore has important implications for the limits of chemical complexity and detection in the ISM. Additionally, tEME is believed to be produced by the same chemical reactions, summarized in Equation 1, that produce dimethyl ether, a molecule detected in numerous environments in the ISM \footnote{R = CH$_3$ for dimethyl ether formation and CH$_3$CH$_2$ for tEME}. Therefore, tEME is the next logical progression in ether synthesis from dimethyl ether. If confirmed, the detection of tEME would represent a significant advance in our understanding of complex molecule formation. 

\begin{align*}
\text{R}\text{OH} + \text{H}_3^+ &\rightarrow \text{R}\text{OH}_2^+ +\text{H}_2       \\
\text{R}\text{OH}_2^+ + \text{CH}_3\text{OH} &\rightarrow \text{CH}_3\text{ORH}^+  +\text{H}_2\text{O}   \tag{1}    \\
\text{CH}_3\text{ORH}^+ + e^- &\rightarrow \text{CH}_3\text{OR} + \text{H}
\end{align*}

In this work, we attempted and failed to confirm the detection of tEME toward W51 e1/e2 using the 12 m Telescope of the Arizona Radio Observatory (ARO) in the 2 mm and 3 mm atmospheric windows, and further report on an extensive search for this species toward the high mass star forming region Sgr B2(N-LMH) with the GBT. We additionally reanalyzed the original detection in the context of the \citet{Snyder1} criteria and show that the reported column density and temperature of \citet{FuchsSpace} are not reproducible based on their observations. Furthermore, no transitions of tEME were conclusively observed toward either W51 e1/e2 or Sgr B2(N-LMH) in the present ARO and GBT data. Our work therefore calls into question the initial detection of tEME toward W51 e1/e2.

%%%%%%%%%%%%%%%%%%%%%%%% Results and Discussion %%%%%%%%%%%%%%%%%%%%%%%%%%%%%
\section{Observations}
\begin{figure}[h!]
\centering
\includegraphics[scale=.35]{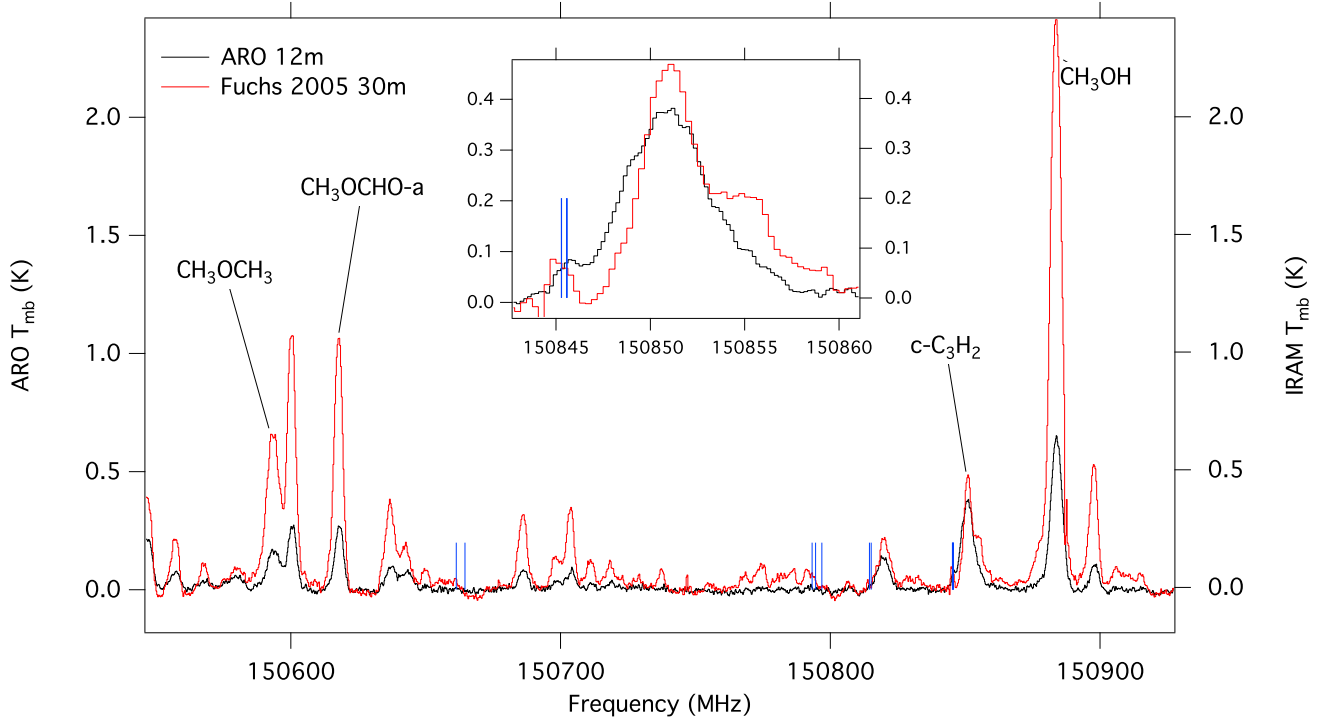}
\vspace{-1em}
\caption{A comparison of the previous IRAM 30 m data and the current ARO 12 m data. All tEME transition frequencies are noted as blue vertical lines of uniform height. The inset shows the tEME 20$_{0,20}$ -- 19$_{1,19}$ transition multiplet. ARO data is converted to T$_{mb}$ assuming the 5\arcsec source size of \citet{FuchsSpace}} and $\eta_m$ = 0.75.
\label{tEME_Comparison}
\end{figure}

\begin{figure}[h!]
\centering
\includegraphics[scale=.90]{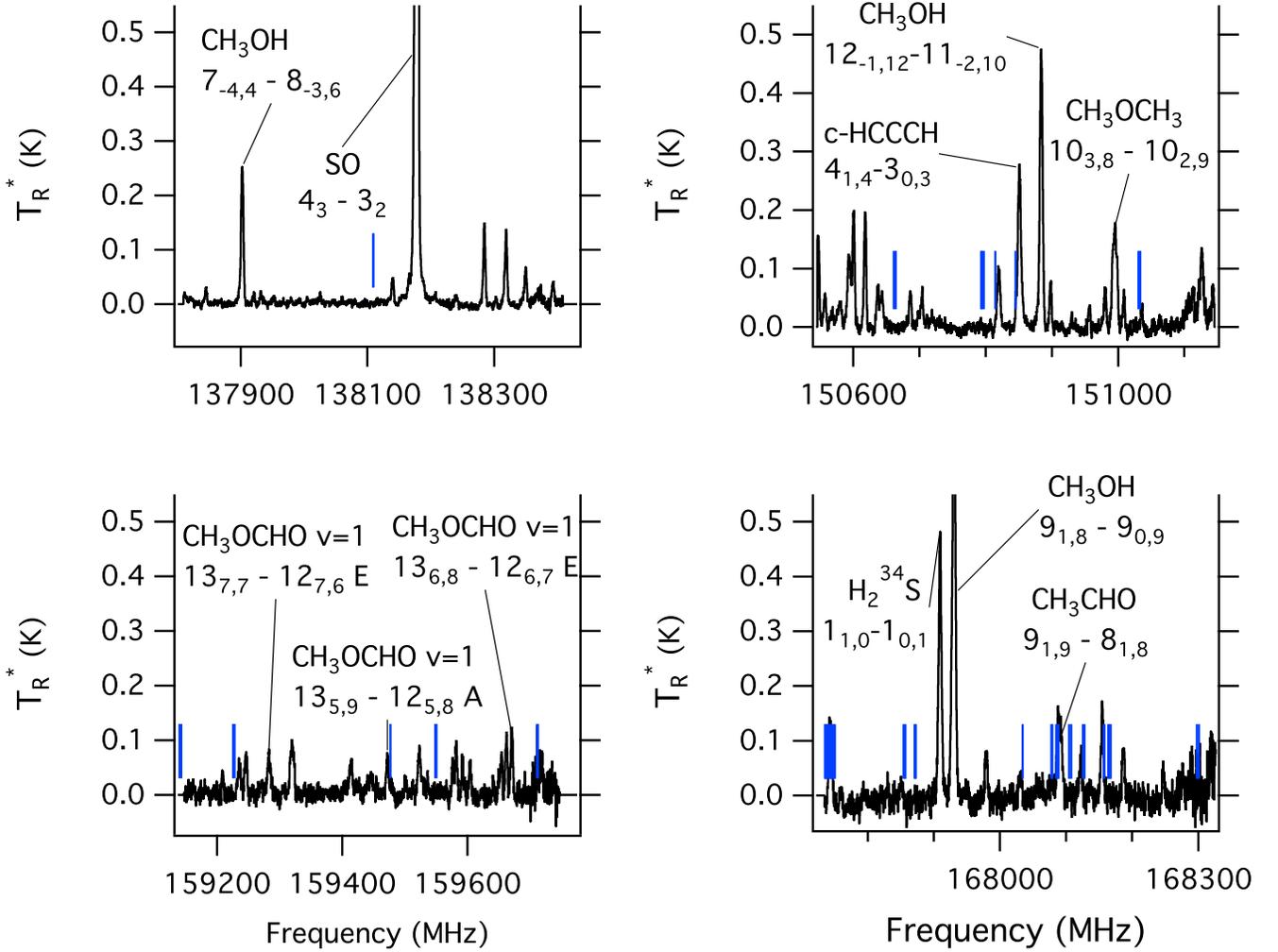}
\vspace{-1em}
\caption{The 2 mm spectral coverage of the ARO observations toward W51 e1/e2. Frequencies are given assuming an LSR velocity of 55 km/s. Molecular transitions are labeled for context. tEME transitions are marked by vertical blue lines of uniform height.}
\label{ARO_Full1}
\end{figure}

\begin{figure}[h!]
\centering
\includegraphics[scale=.90]{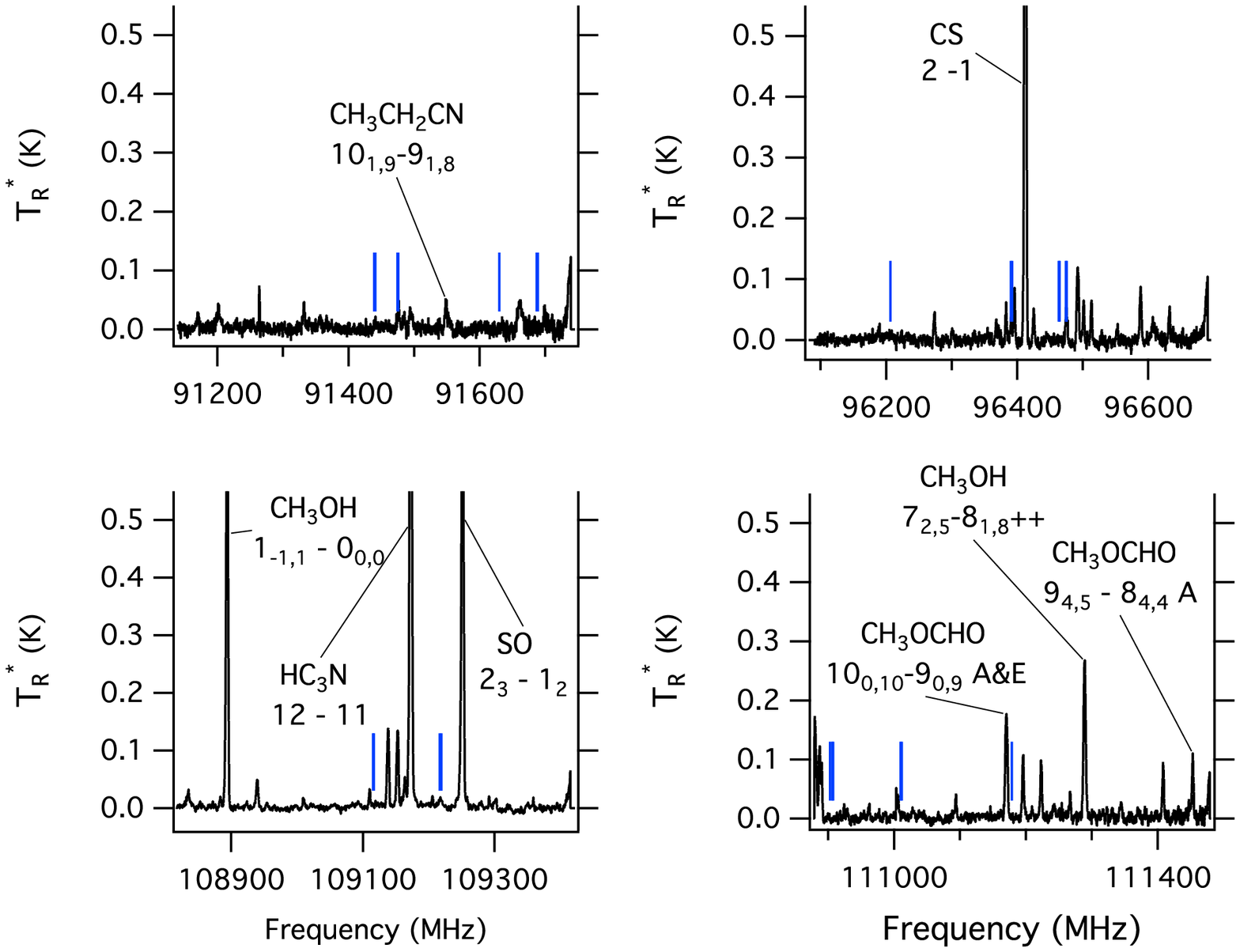}
\vspace{-1em}
\caption{The 3 mm spectral coverage of the ARO observations toward W51 e1/e2. Frequencies are given assuming an LSR velocity of 55 km/s. Molecular transitions are labeled for context. tEME transitions are marked by vertical blue lines of uniform height.}
\label{ARO_Full2}
\end{figure}

The observations using the ARO 12 m telescope, located on Kitt Peak,  were conducted during the period of October 2006 to April 2007. The receivers used were dual-channel, cooled SIS mixers, operated in single-sideband mode with at least 20 dB of image rejection. The back ends used were (1) 256-channel filter banks with 500 kHz and 1 MHz resolution, and (2) a millimeter autocorrelator in the 390.5 kHz resolution mode. All spectrometers were configured in parallel mode to accommodate both receiver channels. The temperature scale, T$^*_R$ , was determined by the chopper-wheel method, corrected for forward spillover losses. Conversion to radiation temperature T$_R$ is then T$_R$ = T$_R^*$/$\eta_c$, where $\eta_c$ is the corrected beam efficiency. Twelve new transitions of tEME covering the range 91 GHz to 168 GHz were studied; over this frequency range, the beam size was 73$\arcsec$ to 38$\arcsec$  and the beam efficiency varied from 0.9 to 0.7. A comparison of the present observations and those from \citet{FuchsSpace} is given in Figure \ref{tEME_Comparison}. A key concern is that the larger beam size of the ARO 12 m telescope may result in beam dilution of potential tEME flux. To assess this possibility, observational frequencies were chosen to partially overlap with those from \citet{FuchsSpace}.   From Figure \ref{tEME_Comparison}, it is likely that both observations sample similar regions, however, the 12 m ARO beam weights more heavily to larger spatial scales than does the IRAM 30 m. The fact that the feature at 150845 MHz attributed by \citet{FuchsSpace} to the 20$_{0,20}$ - 19$_{1,19}$ transition of tEME is slightly stronger in the ARO data indicates a non-compact source size. The complete observations from the ARO 12 m are shown in Figure \ref{ARO_Full1} and \ref{ARO_Full2}.

The observations of Sgr B2 (N) were taken using the National Radio Astronomy Observatory (NRAO) Robert C. Byrd 100 m Green Bank Telescope as part of the \textbf{PR}ebiotic \textbf{I}nterstellar \textbf{MO}lecular \textbf{S}urvey (PRIMOS). Observations began in 2008, and are continually updated.\footnote{The PRIMOS data set is available at \textless {\url{http://www.cv.nrao.edu/~aremijan/PRIMOS/}}\textgreater}  These observations provide nearly continuous high-sensitivity, high-resolution data from 1 GHz to 50 GHz of the Sgr B2(N-LMH) region ($\alpha$[J2000] = 17$^h$47$^m$19.8$^s$, $\delta$[J2000] = -28$^\circ$22$\arcmin$17$\arcsec$). A complete description of the PRIMOS observations can be found in \citet{Neill}. Two tEME transitions at 25.3 GHz and 30.5 GHz were fortuitously covered while searching for other molecules; the telescope beamwidths were $\sim$30$\arcsec$ and $\sim$25$\arcsec$, with corresponding beam efficiencies of 0.7 and 0.6, at those frequencies, respectively.

%%%%%%%%%%%%%%%%%%%%%%%% Results and Discussion %%%%%%%%%%%%%%%%%%%%%%%%%%%%%
\section{Results and Discussion}

\begin{figure}[h!]
\centering
\includegraphics[scale=.3]{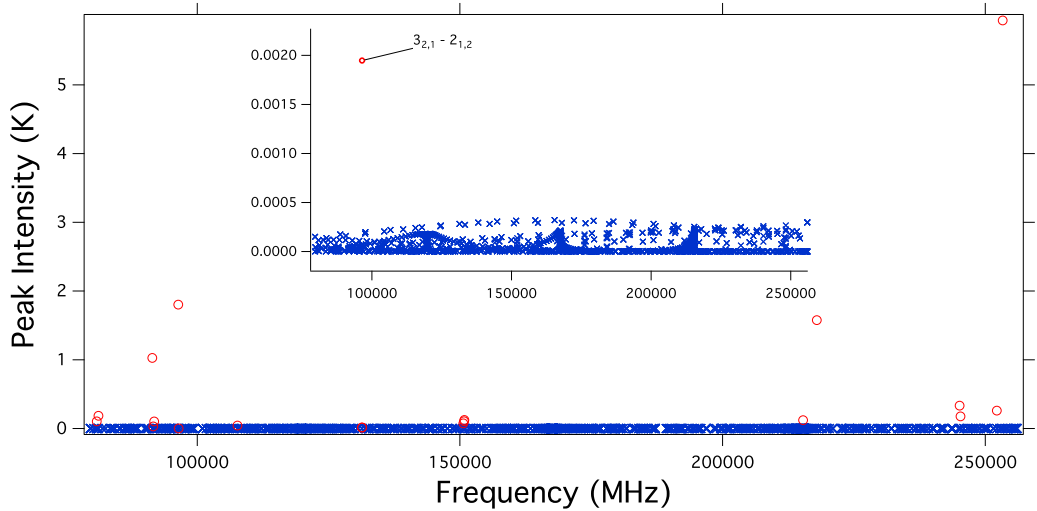}
\vspace{-1em}
\caption{The computed LTE peak antenna temperature of individual tEME transitions based on Equation 2 assuming a column density of 2$\times$10$^{14}$ cm$^{-2}$, velocity width of 3 km s$^{-1}$, and an excitation temperature of 70 K (Blue) plotted with the reported peak antenna temperature from \citet{FuchsSpace} (Red). An enlarged view showing the weakest reported transition from \citet{FuchsSpace} is shown in the inset.}
\label{Fuchs_Int}
\end{figure}

\subsection{Analysis of Previous tEME Observations}
Following the criteria of \citet{Snyder1}, we begin by attempting to verify the previously-reported detection of tEME toward W51 e2. We first consider the possibility that tEME is not well described by an LTE model. While experimental collisional cross-sections are not available, a rough collisional cross section based on molecular geometry and Van der Waals radii gives critical densities of order $\sim$ 10$^3$  - 10$^4$ cm$^{-3}$. Reported densities toward W51 e2 are 10$^3$ - 10$^7$ cm$^{-3}$ \citep{W51Density, W51Density2}, suggesting that tEME transitions should be well described by an LTE approximation. This is supported by the observation that emission from many large species toward W51 is well described by LTE \citep{Kalenskii}. We therefore conclude that an LTE model is appropriate.

A simple first test is to compare the expected local thermodynamic equilibrium (LTE) antenna temperatures with the reported intensity of tEME transitions. From \citet{TonyFormulas}, the LTE antenna temperature is related to column density and temperature by Equation 2, where $E_U$ is the upper state energy of the transition (K), $Q_r$ is the rotational partition function, $\nu$ is the transition frequency (MHz), $S$ is the line strength, $\mu$ is the dipole moment of the molecule (Debye), $\Delta T_{mb}\Delta V$ is the peak observed intensity (mK) times the full width half max (FWHM) of the line (km s$^{-1}$), B is the beam dilution factor, $\Theta_S$ is the source size, $\Theta_B$ is the beam size, and $\eta_B$ is the beam efficiency of the telescope at $\nu$. That is, 

\begin{equation*} 
\label{TonyColumn}
N_T = (1.8\times 10^{14}) \frac{Q_r e^{\frac{E_u}{T_{ex}}}}{\text{B} \nu S\mu^2}\times\frac{\Delta T_{mb}\Delta V}{\eta_B \Bigg(1-\frac {e^{\frac{(4.8\times10^{-5})\nu}{T_{ex}}}-1}{e^{\frac{(4.8\times10^{-5})\nu}{T_{bg}}}-1} \Bigg)}, \hspace{0.1in} B = \frac{\Theta_S^2}{\Theta_B^2 + \Theta_S^2}	\tag{2}\end{equation*}

The transition strengths, frequencies, upper state energies, as well as the rotational partition function (Q = 2027.617 $\times$ T$^{3/2}$) and dipole moment ($\mu_a$ = 0.146 D \& $\mu_b$ = 1.165 D) are taken from \cite{FuchsLab}.  While the assumed velocity width is not explicitly given, from Figure 4 of \citet{FuchsSpace} a FWHM of 1.4 MHz at 150.8 GHz, or 2.7 km s$^{-1}$, may be inferred, in good agreement with previous observations toward W51 e2 \citep[e.g.][]{Remijan1}.  Using the reported column density of 2$\times$10$^{14}$ cm$^{-2}$ and a  rotational temperature of 70 K from \citet{FuchsSpace}, as well as a velocity width of 3 km s$^{-1}$, $\eta_B$ = 1,  a beam filling factor  of B = 1, and background temperature of T$_{bg}$ = 2.7 K, the peak intensity values are calculated in the T$_{mb}$ scale using Equation 2 and plotted (blue crosses) against their corresponding observed values (red circles) from \citet{FuchsSpace} in Figure \ref{Fuchs_Int}. The complete list of parameters used and calculated integrated intensities is given in Table \ref{parameters}.

%, indeed this trend appears to hold for large organic molecules toward many hot cores \citep[e.g.][]{HerschelOrionFullBand}

It is immediately apparent the reported transitions  from \citet{FuchsSpace} do not match their predicted values. Indeed, \textit{every reported transition should have a peak intensity at least an order of magnitude less than its reported value}. As shown in Table \ref{parameters}, the discrepancy is 2 - 4 orders of magnitude for most transitions. In order to be considered valid,  the observed intensity of all transitions should match their predicted values, in accordance with criteria 3 from \citet{Snyder1}.  Accounting for the possibility that the five tEME spin components are blended into a single peak, the greatest peak intensity observed at the column density and temperature reported by \citet{FuchsSpace} would be the 20$_{0,20}$ -- 19$_{1,19}$ transition with a peak intensity of 1.4 mK, well below the previously reported intensity of \citet{FuchsSpace} as well as the RMS of both the present and previous observations.  Reexamining the data as a whole, we performed an iterative least-squares fit of the data used in Fit II of \citet{FuchsSpace} used to determine the reported column density. This yields a best fit column density of  6$\times$10$^{16}$ cm$^{-2}$. We therefore conclude that the column density of 2$\times$10$^{14}$ cm$^{-2}$ derived by \citet{FuchsSpace} is not valid.

A probable explanation for the reported transitions from \citet{FuchsSpace} is interference from coincident transitions of other species.  W51 e1/e2 is a rich molecular source and from the present observations of W51 e1/e2, on average there is a transition with peak intensity $\geq$ 25 mK every 6.3 MHz and a transition with peak intensity $\geq$ 15 mK every 3.2 MHz.  For transitions near the noise level, this means that there is a strong probability that there will be a coincident transition within twice the FWHM that may be falsely attributed to the new molecule. \citet{FuchsSpace} note that of their observed transitions, only two are free of any interfering transitions. This however is based only on comparison with previously detected species and does not account for the possibility of interference from previously unidentified transitions. 

Examining the reported transition frequencies, the difference in the observed and laboratory frequencies varies from -2.0 MHz to 1.46 MHz with a root mean squared value of 926 kHz and a standard deviation of 896 kHz. As these values span a wide range of positive and negative velocity offsets, this cannot be attributed to a systematic difference in the velocity of a single carrier relative to the reported LSR velocity of W51 e2. The laboratory measurements from \citet{FuchsLab} have uncertainties on the order of tens of kHz, thus this also cannot be attributed to uncertainty in the laboratory frequencies. A possible explanation is low spectral resolution. The previously reported astronomical observations have a resolution that varies from 0.3 MHz to 1.25 MHz. Many of the observed transitions have an observed minus calculated value at or below some or all of these spectral resolutions however, because the resolution of the individual spectra used to calculate these values is not specified, it is impossible to evaluate this possibility for many transitions. It can however be noted that four ($\sim$ 21 \%) of the transitions have an observed frequency that differs from its laboratory measurement by $\geq$ 1.25 MHz and are therefore likely not associated with tEME emission. 

\subsection{Analysis of ARO Observations}
\begin{figure}[h!]
\centering
\includegraphics[scale=.75]{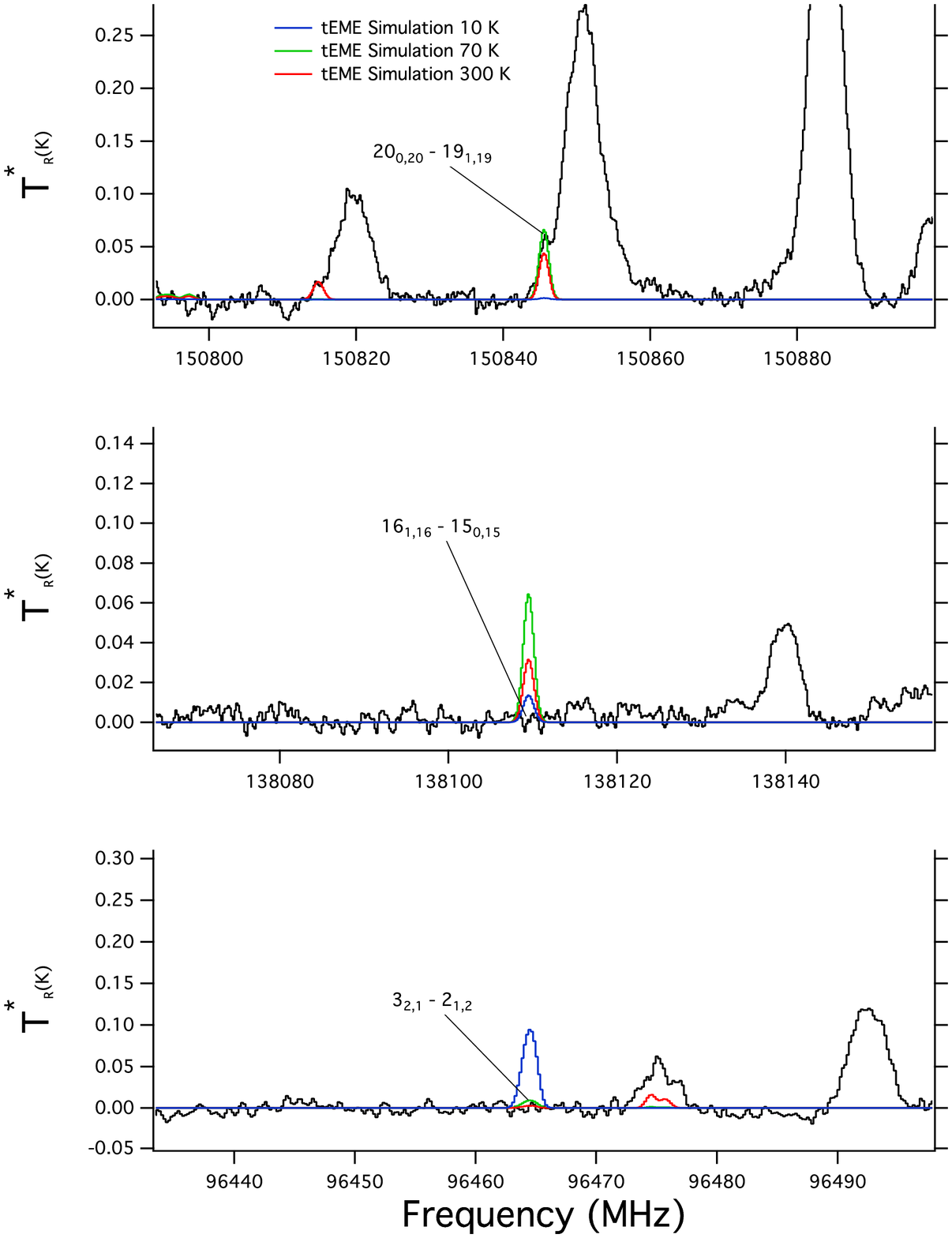}
\vspace{-1em}
\caption{200 km s$^{-1}$ windows of the ARO observations of potential tEME transitions toward W51 e1/e2 (Black). A simulation of tEME at 10 K (Blue) 150 K (Green) and 300 K (Red), assuming $\Delta$V = 3 km s$^{-1}$, $\eta_B$ = 1, B = 1, N$_T$ is the best fit column density derived for each temperature. Clearly the 3$_{2,1}$ - 2$_{1,2}$ and 16$_{1,16}$ - 15$_{0,15}$ are inconsistent with the 20$_{0,20}$ -- 19$_{1,19}$ transition.}
\label{FuchsFail}
\end{figure}

An alternative approach is to examine all tEME transitions covered by the present ARO observations. As a starting point, we assume a column density of 1.3$\times$10$^{16}$ cm$^{-2}$ such that the emission at the 20$_{0,20}$ -- 19$_{1,19}$ tEME transition is reproduced for an excitation temperature of 70 K.  A simulation can be made of the resulting tEME line intensities, as shown in Figure \ref{FuchsFail}. It is clear in this modeling that several transitions with predicted peak intensities well above the RMS of the observations are clearly absent. To satisfy criteria 4 of \cite{Snyder1}, there should be no absent transitions.  In fact, an excitation temperature of 70 K cannot satisfy this criteria unless the column density is sufficiently low that all observed transitions have peak intensities below the RMS of the observations. Considering other excitation temperatures (10 K - 300 K) and column densities (1$\times$10$^{12}$ cm$^{-2}$ - 1$\times$10$^{16}$ cm$^{-2}$) does not improve the situation. It becomes apparent that, in order not to have ``missing" lines, the tEME column density must be sufficiently low that all observed transitions are below the RMS of the observations from \cite{FuchsSpace} as well as the present observations, and thus not detectable in either observation.

An additional concern is the effect of beam dilution. Fuchs et al. assume a source size of 5$\arcsec$. At 145 GHz, the ARO beam is $\sim$ 43$\arcsec$, corresponding to a beam dilution factor 6.67 times higher than that of the IRAM 30 m at the same frequency. If this source size is correct, the present ARO observations would be up to a factor of 6.67 less sensitive. However, examining the only transition covered by both observatories, the 20$_{0,20}$ -- 19$_{1,19}$ tEME transition and 150845 MHz (Figure \ref{tEME_Comparison}), after comparing both ARO and IRAM 30 m observations in the T$_{mb}$ scale, assuming a 5\arcsec source size, it is clear that the flux observed at this frequency does not decrease when observed with a larger beam, indicating that it cannot arise from a compact source. Therefore a beam dilution factor of 6.67 cannot apply. Furthermore, the column density from Fit II of \cite{FuchsSpace} would still produce transitions clearly visible in the ARO observations.

With no reliably identified tEME transitions, we determine an upper limit to the column density towards W51 e1/e2 using the current observations. As dimethyl ether and tEME are thought to form from similar processes, it is plausible to assume that they should have similar excitation conditions in a source. From \citet{Kalenskii}, the derived rotational temperature of dimethyl ether towards W51 e1/e2 is 85 K. The strongest tEME transition in the current observations at 85 K that has no obvious interfering transitions is the 12$_{0,12}$ -- 11$_{0,11}$ transition. This transition is not detected, but the RMS at this frequency can be used to determine an upper limit. Using Equation 2 and a velocity width of 3 km s$^{-1}$, an upper limit of $\leq$ 1$\times$10$^{15}$ cm$^{-2}$ can be derived for tEME, assuming all five components are blended into a single transition. 

Finally we assess the possibility of detecting tEME in Sagittarius B2 (N-LMH). Using data from the PRIMOS project toward Sgr B2 (N-LMH), we have searched for possible tEME transitions. Several peaks coincident with tEME transitions were located, however several tEME transitions of similar predicted intensity show no emission, indicating that these features are simply coincidental.  We therefore use the RMS at the strongest predicted transition to set an upper limit. Molecules detected toward Sgr B2 (N-LMH) show a wide range of excitation temperatures. No tEME transitions are detected, making it impossible to determine an excitation temperature. We therefore compute the upper limit at 10 K and 85 K. At 10 K, the strongest transition with no interfering features is the 4$_{1,3}$ -- 4$_{0,4}$ transition, while at 85 K the strongest feature would be the 9$_{1,8}$ -- 9$_{0,9}$ transition. The RMS at each transition frequency in the PRIMOS data is 4.5 mK and 11 mK, respectively. Using Equation 2 and assuming a beam efficiency of 0.8, the molecular parameters given in Table \ref{parameters}, and a velocity width of 13 km s$^{-1}$, the upper limits for the column density of tEME towards Sgr B2 (N-LMH) are $\leq$ 2.1 $\times$ 10$^{15}$ cm$^{-2}$ and $\leq$ 1.7 $\times$ 10$^{16}$ cm$^{-2}$, respectively. 

\clearpage

\begin{deluxetable}{c c c c c c c}
\tablecolumns{7}
\tabletypesize{\footnotesize}
\tablecaption{Observed and calculated intensity of tEME Transitions.\tablenotemark{a}}
\tablewidth{0pt}
\tablehead{
\colhead{Transition}										&	\colhead{$\nu$}		&	\colhead{S$_{ij}$}				&	\colhead{E$_u$}		&	\colhead{N$_{Lines}$}	&	\colhead{Observed	 $\int T_{mb}$d$v$}\tablenotemark{b}	&	\colhead{Calculated $\int T_{mb}$d$v$\tablenotemark{c}} \\
\colhead{$J_{K_a,K_c}^{\prime} - J_{K_a,K_c}^{\prime\prime}$}	&	\colhead{(MHz)}		&	\colhead{}						& 	\colhead{(K)}			&	\colhead{}				&	\colhead{ K km s$^{-1}$}				&	\colhead{ K km s$^{-1}\times$10$^3$}
}	
\startdata
11$_{2,10}$ -- 11$_{1,11}$								&  80881.71 - 80883.6		&	5.195							&		30.1				&		5				&	0.09							&		0.3695\\
24$_{1,23}$ -- 24$_{0,24}$								&  81198.23 - 81199.22 		&	10.268							&		118.5			&		5				&	0.16							&		0.2073\\
35$_{3,32}$ -- 35$_{2,33}$								&  91439.39 - 91441.26		&	27.608							&		255.2			&		5				&	1.0							&		0.0891\\
34$_{2,32}$ -- 34$_{1,33}$								& 91630.26 - 91631.17		&	21.850							&		237.1			&		5				&	0.03							&		0.09149\\
37$_{3,34}$ -- 37$_{2,35}$								& 91811.70 - 91813.40		&	29.637							&		283.7			&		5				&	0.1							&		0.06386\\
29$_{3,26}$ -- 29$_{2,27}$								& 96390.20 - 96392.55		&	20.487							&		179.1			&		5				&	1.85							&		0.2067\\
3$_{2,1}$ -- 2$_{1,2}$									& 96463.73 - 96464.85		&1.545 - 1.639 							&		6.9				&		5				&	0.002						&		0.1936\tablenotemark{f}\\
22$_{3,19}$ -- 22$_{2,20}$								& 107655.40 - 107658.06		& 	13.199							&		108.3			&		5				&	0.05							&		0.4089\\
7$_{2,5}$ -- 6$_{1,6}$									& 131349.80 - 131351.62		&	2.306							&		15.4				&		5				&	0.02							&		0.3284\\
15$_{1,15}$ -- 14$_{0,4}$									& 131372.62 - 131373.11		&	9.858							&		46.7				&		5				&	0.02							&		0.8986\\
34$_{4,30}$ -- 34$_{3,31}$								& 150661.35 - 150664.55		&	20.064							&		248.7			&		5				&	0.12							&		0.1170\\
13$_{6,x}$ -- 14$_{5,y}$\tablenotemark{d}						& 150793.24 - 150797.40		&	1.289							&		76.6				&		10				&	0.17							&		0.088\\
20$_{0,20}$ -- 19$_{1,19}$								& 150845.28 - 150845.58		&	14.283							&		80.4				&		5				&	0.20							&		0.9231\\
19$_{5,z}$ -- 19$_{4,15}$\tablenotemark{e}					& 215324.99 - 215327.56		& 3.487 - 9.519	 						&	 	102.2			&		6				&	0.28							&		0.6436\tablenotemark{f}\\
28$_{0,28}$ -- 27$_{1,27}$								& 217940.65 - 217940.76		&	22.586							&		154.7			&		5				&	3.66							&		0.7299\\
16$_{3,14}$ -- 15$_{2,13}$								& 245103.55 - 245106.43		&	4.994							&		62.9				&		5				&	0.46							&		0.6735\\
31$_{1,31}$ -- 30$_{0,30}$								& 245274.09 - 245274.22		&	25.712							&		188.8			&		5				&	0.87							&		0.5748\\
17$_{3,15}$ -- 16$_{2,14}$								& 252188.29 - 252191.14		&	5.191							&		69.5				&		5				&	0.70							&		0.6556\\
28$_{2,27}$ -- 27$_{1,26}$								& 253307.71 - 253308.94		&	12.415							&		161.0			&		5				&	16.01						&		0.4264\\
\hline
4$_{1,3}$ -- 4$_{0,4}$\tablenotemark{g}						&   25335.53 - 25336.17		&	4.386							&		5.1				&		5				&								&			\\
9$_{1,8}$ -- 9$_{0 ,9}$										&   30561.87 - 30562.54		&	8.335							&		18.2				&		5				&								&			\\
\enddata
\tablenotetext{a}{Computed integrated intensities are assuming T$_{ex}$ = 70 K, N$_T$ =   2$\times$10$^{14}$ cm$^{-2}$, and $\Delta$V = 3 km s$^{-1}$, B = 1}
\tablenotetext{b}{Observed integrated intensities are taken from \citet{FuchsSpace}}
\tablenotetext{c}{Computed integrated intensities are for a single transition. The maximum observable integrated intensity can be obtained by $\int T_{mb}$d$v_{max}$ = $\int T_{mb}$d$v\times$N$_{Lines}$}
\tablenotetext{d}{ \textit{x}-\textit{y} deontes 7-9, 7-10, 8-9, or 8-10.}
\tablenotetext{e}{\textit{z} denotes either 14 or 15.}
\tablenotetext{f}{Computed using the average value of S$_{ij}$}
\tablenotetext{g}{Parameters used to derive upper limits for PRIMOS data}
\label{parameters}
\end{deluxetable}	

\pagebreak

\section{Conclusions}
Rigorous application of the criteria for detection of a new molecule as outlined in \citet{Snyder1} has again been applied to a claimed detection. As in the case of dihydroxyacetone \citet{Apponi1} and glycine \cite{Snyder1},  these criteria underline the need for a thorough analysis when evaluating the possible detection of new molecules. In the present case, analysis of the previously reported detection of tEME (\citealt{FuchsSpace}) calls into question the original detection. Both the LSR velocities and LTE intensities reported in \citet{FuchsSpace} are shown to be inconsistent with the reported column density and temperature, casting doubt on the claimed detection of tEME. Based on previous observations of W51 e1/e2, we instead derive an upper limit five times higher than the previously reported value. We also derive similar upper limits toward Sgr B2 (N-LMH). 
 
P.B.C \& G.A.B. gratefully acknowledge funding from the NASA Astrophysics Research and Analysis and Herschel Guaranteed Time Observer programs. B.A.M. gratefully acknowledges funding by an NSF Graduate Research Fellowship. The Arizona Radio Observatory is operated by Steward Observatory, University of Arizona, with partial support through the NSF University Radio Observatories program (URO: AST-1140030). The National Radio Astronomy Observatory is a facility of the National Science Foundation operated under cooperative agreement by Associated Universities, Inc.

%Large asymmetric molecules, particularly those that posses splitting from internal motions of the molecule have a large number of transitions in the millimeter regime, making the possibility of coincident transitions extremely high in molecule-rich sources.

\clearpage

\bibliographystyle{apj}
\bibliography{tEME}

\end{document}